\documentclass[10pt,letterpaper]{article}
\usepackage[top=0.85in,left=1in,footskip=0.75in]{geometry}

\raggedright
\usepackage{amsmath,amssymb}

\usepackage{changepage}

\usepackage[utf8x]{inputenc}

\usepackage{textcomp,marvosym}

\usepackage{cite}

\usepackage{nameref,hyperref}

\usepackage{microtype}
\DisableLigatures[f]{encoding = *, family = * }

\usepackage[table]{xcolor}

\usepackage{array}

\newcolumntype{+}{!{\vrule width 2pt}}

\newlength\savedwidth

\usepackage{setspace} 
\usepackage[aboveskip=1pt,labelfont=bf,labelsep=period,justification=raggedright,singlelinecheck=off]{caption}
\makeatletter
\renewcommand{\@biblabel}[1]{\quad#1.}
\makeatother

\usepackage{lastpage,fancyhdr,graphicx}
\usepackage{epstopdf}
\fancyheadoffset[L]{1in}
\begin{document}
\vspace*{0.2in}

\begin{flushleft}
{\Large
\textbf\newline{Walking speeds for short distances and turning in lower-limb amputees: a study in low-cost prosthesis users} % Please use "sentence case" for title and headings (capitalize only the first word in a title (or heading), the first word in a subtitle (or subheading), and any proper nouns).
}
\newline
\\
Nidhi Seethapathi\textsuperscript{1,3*},
Anil Kumar Jain \textsuperscript{2},
Manoj Srinivasan\textsuperscript{1},
\\
\bigskip
\textbf{1} Mechanical and Aerospace Engineering, The Ohio State University, Columbus OH 43210, USA
\\
\textbf{2} Santokba Durlabhji Memorial Hospital, Jaipur, Rajasthan 302015, India
\\
\textbf{3} Department of Bioengineering, University of Pennsylvania, Philadelphia PA 19104, USA
\\
\bigskip

% Insert additional author notes using the symbols described below. Insert symbol callouts after author names as necessary.
% 
% Remove or comment out the author notes below if they aren't used.
%
% Primary Equal Contribution Note

% Additional Equal Contribution Note
% Also use this double-dagger symbol for special authorship notes, such as senior authorship.

% \textcurrency Current Address: Dept/Program/Center, Institution Name, City, State, Country % change symbol to "\textcurrency a" if more than one current address note
% \textcurrency b Insert second current address 
% \textcurrency c Insert third current address

% \dag Deceased

% Group/Consortium Author Note
% \textpilcrow Membership list can be found in the Acknowledgments section.

% Use the asterisk to denote corresponding authorship and provide email address in note below.
* snidhi@seas.upenn.edu

\end{flushleft}
\section*{Abstract}
Preferred walking speed is a widely-used performance measure for people with mobility issues, but is usually measured in straight line walking for fixed distances or durations, and without explicitly accounting for turning. However, daily walking involves walking for bouts of different distances and walking with turning. Here, we  studied walking for short distances and walking in circles in unilateral lower-limb amputees wearing an above or below-knee passive prosthesis, specifically, a Jaipur foot prosthesis. Analogous to earlier results in non-amputees, we found that their preferred walking speeds are lower  for short distances and  lower for circles of smaller radii.  Using inverse optimization, we estimated the cost of changing speeds and turning such that  the observed preferred walking speeds in our experiments minimizes the total energy cost. The inferred costs of changing speeds and turning were much larger than for young non-amputees. These findings could inform prosthesis design and rehabilitation therapy to better assist changing speeds and turning tasks in amputee walking. Further, measuring the preferred speed for a range of distances and radii is a more robust subject-specific measure of walking performance. % TODO: add jaipur foot and widely used and under studied.

% Please keep the Author Summary between 150 and 200 words
% Use first person. PLOS ONE authors please skip this step. 
% Author Summary not valid for PLOS ONE submissions.   
\section*{Author summary}
Despite being one of the most widely used prosthesis in the world, much is unknown about amputee walking behavior while wearing the Jaipur foot prosthesis. Here, we measure their preferred walking speed while walking for a range of distances and while walking in circles of different radii. Using a within subjects design, we find that amputee subjects walk slower on average for short distances and slower on average on smaller curves. % TO DO

% \linenumbers

% Use "Eq" instead of "Equation" for equation citations.
\section*{Introduction}
Overground walking speed is commonly used to quantify a human subject's mobility improvement after being fit with a new prosthetic leg or after undergoing physical therapy or rehabilitation from stroke, other injury, or movement disorder \cite{bohannon1997comfortable,boonstra1993walking}. Here, we study the preferred walking speeds for subjects with amputation while walking for short distances or while turning, showing that these change systematically for shorter distances or higher curvatures of turning. 

Walking speeds are estimated using a variety of tests in the lab, for instance, using the 6 minute walk test \cite{harada1999mobility} or the 10 m walk test \cite{amatachaya2014concurrent}, in which the tests are framed as cardiovascular endurance tasks with the subjects being asked to ``walk as far as possible in the given duration''. An alternative walking speed measure is obtained by having subjects walk short distances such as 3 m \cite{graham2008assessing}, 4 m \cite{peters2013assessing}, 5 m \cite{wilson2013utilization}, and 15 m \cite{dobkin2006short,dickstein2008rehabilitation,bohannon1997comfortable} in a sub-maximal `comfortable' or `natural' manner. Here, we focus on the latter version, where amputee subjects walk naturally at their `preferred walking speeds' for relatively short distances. In healthy adults with no movement disorders, the preferred speed for walking in a straight line was recently shown to be  distance-dependent \cite{seethapathi2015metabolic}: the speed is systematically lower for shorter distances. This distance-dependence can be explained by the larger energetic cost of speeding up and slowing down for shorter distances \cite{seethapathi2015metabolic}. Here, we characterize this distance dependence of walking speeds in unilateral amputees wearing the Jaipur foot prosthesis. We propose that this distance dependence of walking speed could be a more complete measure of preferred walking speed (compared to measuring the speed at just one distance). This distance dependence of walking speed is also relevant because a considerable percentage of daily walking occurs in short bouts \cite{orendurff2008humans}, especially in amputees \cite{klute2006prosthetic,shell2017effects}.

Effective mobility also requires ability to walk with turning \cite{gailey2002amputee}.  Indeed, for subjects in one previous study, between 8\% and 50\% of all walking steps in daily life involved turning \cite{glaister2007video}. As a way of quantifying turning ability, here, we propose the measurement of preferred speeds while walking in circles of different radii. Here, we characterize such circle walking speeds in unilateral amputees. In non-amputee adults, the tangential speed of walking was recently shown to depend on the curvature of the circle walked: slower walking for smaller circles or higher curvature \cite{brown2021unified}. This slow-down was explained by the increased energetic cost of walking with turning \cite{brown2021unified}. Some curved walking interventions have been considered for other populations \cite{lowry2012contributions,godi2010curved,hess2010walking}. The mechanics of amputees walking in a circle has been studied in some detail \cite{segal2011comparison,ventura2011compensatory,shell2017effects}, but the radius dependence of speeds has not previously been characterized.

Metabolic energy optimality has been used to make predictions for a number of aspects of overground and treadmill walking behaviors  in non-amputee adults \cite{zarrugh1974optimization,srinivasan2009optimal,donelan2001mechanical}. 
 Indeed, as noted above, the preference for slower walking speeds for shorter distances and slower walking speeds for smaller radii and higher curvatures in non-amputees have been attributed to optimization of the corresponding energy costs \cite{seethapathi2015metabolic,brown2021unified}. Here, we interpret the measured amputee walking for short distances and in curves from the perspective of energy minimization.
 In an early classic, Ralston \cite{ralston1958energy} showed that both non-amputee and amputee walkers preferred to walk close to their energy-optimal speeds in preferred speed experiments, even though they could walk faster. More recently, there has been related work on prosthesis aimed at reducing energy expenditure in walking \cite{handford2016robotic,collins2005controlled,herr2012bionic,beck2017reduced,quesada2016increasing} and understanding locomotor adaptation in populations with amputations or other movement disorders from an energetics perspective \cite{hansen2010net,finley2013learning,fey2012optimization,handford2016robotic,collins2005controlled,quesada2016increasing}. Amputees walking with passive prosthetic legs usually have a higher steady state metabolic cost \cite{waters1976energy} compared to non-amputees, but recent work finds no significant difference in some amputee sub-populations \cite{esposito2014does}. In either case, it is not known whether amputees have higher energy costs of speeding up, slowing down, and turning when walking, compared to non-amputee walkers. By viewing preferred walking speeds in amputees through the lens of energy minimization, these short distance and circle walking tasks may provide insight into how their energy cost of changing speeds and the cost of turning may compare to that of non-amputee individuals. 
 
In summary, in this study, we examine the distance-dependence of preferred straight-line walking speeds and the radius-dependence of walking speeds while turning. We study these dependencies in a population of unilateral amputees in India, both above- and below- knee, using the Jaipur foot prosthesis. The Jaipur foot is a low cost prosthesis used widely in the developing world, developed by P. K. Sethi and co-workers  in the 1970s \cite{sethi1978vulcanized}. It was designed with greater mobility in the ankle and subtalar joint for facilitating movements and postures common in India, such as bare foot or shod walking over unpaved uneven terrain and squatting or cross-legged sitting on the floor. It has three main pieces: a heel block and fore-foot-toe block made out of micro-cellular rubber bonded to a wooden ankle block. Tread rubber is used on the undersurface to provide traction and the entire assembly is covered in a skin colored compound to provide cosmesis and water proofing. The Jaipur foot is used in over 22 countries and by hundreds of thousands of amputees, most widely used second only to the SACH foot \cite{howitt2012technologies,arya2008jaipur}. Thus, this study adds to the small number of biomechanical studies on the Jaipur foot \cite{arya1995biomechanical,lenka2010gait,mishra2018performance,mishra2019performance}, which remains an under-studied prosthesis despite being so widely used.

\section*{Materials and methods}
\begin{figure*}[h!]
\begin{center}
 \includegraphics{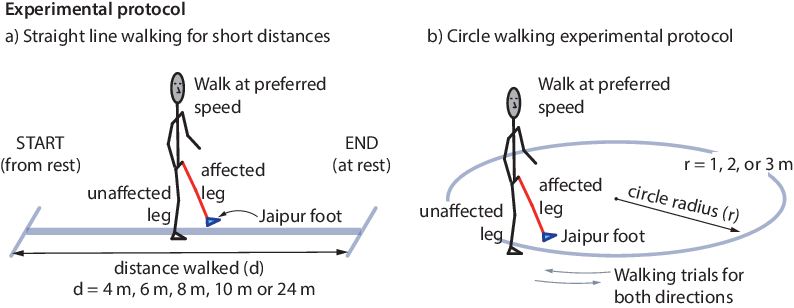}
	\caption{\textbf{Overground walking experiment setup.} We measured the preferred walking speed of walking for unilateral amputees wearing a Jaipur foot passive prosthetic leg in two conditions: a) walking a range of short distances, starting and stopping each bout at rest and b) walking in circles of different radii, both clockwise and anti-clockwise.}
	\label{fig:JFoot1}
\end{center}
\end{figure*}

\paragraph{Subject population.}
The experimental protocol was carried out in India and all subjects participated with informed verbal consent, as approved by the Ohio State University Institution Review Board. Verbal consent was obtained by the experimenter by first explaining the procedures, risks, and benefits in the subjects' native language (Hindi), as approved by the IRB. All subjects ($N = 12$ with 11 male, 1 female, $65.75 \pm 12.6$ kg with prosthesis and shoes, height $1.67 \pm 0.09$ meters and age $39 \pm 14.09$ years, mean $\pm$ s.d.) were unilateral amputees, out of which 7 subjects were above-knee amputees and 5 were below-knee amputees. The right leg was the affected leg for nine subjects and the left for three subjects. See \textit{Supplementary Information (S1 Data file)} and Table \ref{table:Subjects} for individual subject data.  All subjects had a unilateral Jaipur Foot prosthesis \cite{sethi1978vulcanized}, either the above- or below-knee prosthesis, manufactured and fit in the Santokba Durlabhji Memorial Hospital in Jaipur. All walking trials were also conducted at this hospital. Inclusion criteria stipulated that the subjects be able to walk independently without using canes, crutches, hand rails, or assistive devices distinct from their prostheses.  Each subject performed two kinds of walking trials: (i) walking for short distances and (ii) walking in circles, as described below (Fig \ref{fig:JFoot1}). Subjects' walking was video-recorded and time durations to complete a walking task were timed using a stop-watch and verified using the video. The subjects did not carry any additional instrumentation. The subjects walked shod in the reported trials, using their daily foot-wear. % add time spent on feet, prosthesis lengths?

\paragraph{Experiment: Walking for short distances.}
Each subject was instructed to walk in a straight-line for five different short distances $D$: 4 m, 6 m, 8 m, 10 m and 23 m (Fig \ref{fig:JFoot1}a). There were four trials for each distance, resulting in 20 walking trials per subject. Trial order was randomized. Subjects were asked to ``walk the way they usually walk'' and they had to start and end each trial standing still, so they had to speed up from rest and slow down to rest. Average walking speeds were estimated by measuring the time duration $T$ for each trial (starting and ending at rest) and computing $D/T$. Between any two trials, subjects had breaks of durations between 15 seconds to up to a minute. 

\paragraph{Experiment: Walking in circles.}
Subjects were asked to walk in circles of three different radii: 1 m, 2 m and 3 m (Fig \ref{fig:JFoot1}b), completing 5, 4 and 3 laps, respectively, for these radii. For each radius, subjects performed two trials, once with the prosthetic leg inside the perimeter and once with the prosthetic leg outside the perimeter of the circle. Trial order was randomized over the circle radii and walking directions. The average speed was obtained by measuring the total walking duration and averaging over all laps for each trial. Subjects walked with the circle between their two feet, maintaining a non-zero step width, rather than step on the circle with both feet. Between any two trials, subjects had breaks of durations between 15 seconds to up to a minute. 

% \pagebreak
\paragraph{Hypothesis testing.} We tested three primary \textit{a priori} hypotheses described as follows: 
\renewcommand{\labelenumi}{\textbf{\arabic{enumi}.}}
\begin{enumerate}
\item \textbf{Average walking speed is lower for shorter distances.} For this distance-dependence test, we performed two types of tests: (1) We test whether each of the four shorter distance bouts (4, 6, 8, 10 m) are slower on average than the longer distance bout (23 m). This involves four tests, one for each distance. (2) We fit a linear model to the shorter distance bouts (4-10 m) to see if the speed vs distance line has positive slope.

\item  \textbf{Average walking speed is smaller for smaller radii while walking in circles.}  Similarly, for the radius dependence experiment, we performed two types of tests: (1) We test whether walking on each of the three circles (1, 2, 3 m) are slower than walking in a straight line. This results in 3 tests, one for each radius. (2) We fit a linear model to the circle walking speeds to see if the speed vs radius line has positive slope. 

\item  \textbf{Subjects walked slower in circles when the prosthesis leg was on the inside.} To test this hypothesis, we tested whether subjects walked slower in circles when the prosthesis leg was on the inside by pooling across all amputees and all radii.
\end{enumerate}

To test these hypotheses, we used a within-subjects design (repeated measures), where the comparisons are performed between the subjects' own speeds at different distances and different radii. We did not perform across-subjects or across-population comparisons. All hypothesis tests performed were paired (that is, within subject comparison and one-sided). The aforementioned linear models (speed vs distance and speed vs radius) use subject-specific offsets to account for the systematic speed differences between subjects.  We performed the tests non-parameterically using bootstrap resampling with $10^5$ bootstrap samples \cite{perry2017walking,hall1991two,chernick2011bootstrap}; this non-parametric approach allows for data complexity and non-normality due to the heterogeneous subject pool, accounting for correlated behavior for the different distances and radii by a given subject. The number of tests of statistical significance listed above is 10 and we simply used a Bonferroni correction to the $p$ values to control for multiple comparisons. For comparison with the bootstrap procedure, we also provide one-sided t-test-based $p$ values as well, also with Bonferroni correction.  We label any additional tests presented in the Results section as post hoc exploratory analysis.

% We perform hypothesis tests using bootstrap resampling, which 
% 12 subjects: are the subject numbers sufficient for inference?
% what else do we test? exploratory data analysis. above vs below knee, inside vs outside leg

\paragraph{Mathematical model: walking for short distances.}
For short distance walking, we compare the experimentally observed preferred walking speed results to the walking speed predicted by minimizing the total metabolic cost of the walking bout. For simplicity, we assume that people walking a distance $D$ start from rest (specified in experiment), then instantaneously speed up to some speed $v$, continue at that speed for the whole distance and then instantaneously come to rest again. Thus, the total cost of walking the distance includes the cost of accelerating from rest to speed $v$ at the start, walking at constant speed $v$ and then decelerating to rest at the end of the walking bout, given by the following equation (following the approach in \cite{seethapathi2015metabolic}):
\begin{equation}
    E_\mathrm{met} = (a_0 + a_1v + a_2 v^2) \frac{D}{v} + \lambda \left(\frac{1}{\eta_\mathrm{pos}}+\frac{1}{\eta_\mathrm{neg}} \right) \left( \frac{1}{2}mv^2 \right). \label{eq:changingshortdist}
\end{equation}
Here,  $\dot{E} = a_0 + a_1 v + a_2 v^2$ in Wkg$^{-1}$, with $v$ in ms$^{-1}$ models the metabolic rate of walking $\dot{E}$ at a constant speed $v$ for both amputees and non-amputees, with $a_0 = 4.97$, $a_1 = -5.78$, and $a_2 = 5.62$  for above-knee amputees \cite{Jaegers1993},  $a_0 = 3.24$, $a_1 = -2.19$, and $a_2 = 2.89$ for below-knee amputees \cite{Genin2008}, and $a_0 = 2.22$, $a_1 = 0$, and $a_2 = 1.155$ for non-amputees \cite{Jaegers1993}. All of these relations for $\dot{E}$ result in a classical U-shaped relationship between energy cost per unit distance and speed of walking. The quantity $\lambda$ provides a scaling factor between the kinetic energy $mv^2/2$ and the energy cost required to achieve it, with $\eta_\mathrm{pos}  = 0.25$ and $\eta_\mathrm{neg} = 1.2$ are traditional muscle efficiencies for performing positive and negative mechanical work, respectively \cite{srinivasan2010fifteen}. 

The speed that minimizes the short-distance cost of walking $E_\mathrm{met} $ is obtained by differentiating the total energy expression in equation \ref{eq:changingshortdist} and is given by the implicit equation: $\lambda v^3 \left({\eta_\mathrm{pos}}^{-1}+{\eta_\mathrm{neg}}^{-1} \right)/(a_0 - a_2 v^2) = D$. This relation implies that shorter distance bouts should have lower speeds \cite{seethapathi2015metabolic}.

\paragraph{Mathematical model: Walking in circles.} Following the approach in \cite{brown2021unified}, who directly measured the cost for walking in circles for non-amputee subjects, we analogously propose that the metabolic rate of walking in a circle for amputees is $\dot{E} = a_0 + a_1 v + a_2 v^2 + b_2 (v/R)^2$, where $R$ is the circle radius, with $a_{0,1,2}$ values as above. The term $b_2 (v/R)^2$ is the additional cost of turning. The cost per distance is given by: 
\begin{equation}\dot{E}/v = a_0/v + a_1 + a_2 v + b_2 v/R^2, \label{eq:Eturning}
\end{equation} 
which is minimized by the speed $v = \sqrt{a_0/(a_2 + b_2/R^2)}$. Thus, the prediction is that the optimal speed is smaller for smaller radius $R$. 

For both short-distance walking and circle walking, we also determine the speeds that are within 1\% of the minimum energy cost, because the energy landscapes are usually flat and a small change in speed near the minimum energy usually results in a much smaller energy change.

\paragraph{Inverse optimization to estimate the cost of changing speeds and turning.} We did not experimentally measure the energy costs of changing speeds (as in \cite{seethapathi2015metabolic}) or turning (as in \cite{brown2021unified}) in amputees. Instead, we used inverse optimization to estimate the cost coefficients  $\lambda$ and $b_2$ for changing speeds and turning respectively. Inverse optimization is a model-fitting procedure in which parameters governing the energy landscape are obtained so that we obtain the experimentally observed behavior when this energy cost is minimized  \cite{mombaur2010human,liu2005learning,mainprice2015predicting}. Specifically, we determined coefficients $\lambda$ and $b_2$ values such that the speed reduction exhibited by our amputee subjects is predicted well by minimizing the total energy cost of the respective tasks (Eqs \ref{eq:changingshortdist} and \ref{eq:Eturning}). To do this, we performed a series of optimizations for different values of $\lambda$ and $b_2$, and picked that which minimized the summed squared difference between predicted optimal speeds and observed preferred speeds: $\sum \left( v_\textrm{model}-v_\textrm{data} \right)^2 $, summed over all subjects and trials. This fit was performed separately for the above knee and the below knee amputees. 
% To do: Say, we did this for all above knee and below knee amputees together and not for individuals?

% \paragraph{Non-amputee data.} For purely visualization and qualitative comparison purposes in the figures, we use data from our prior non-amputee experiments for short distances \cite{seethapathi2015metabolic} and for walking in circles \cite{brown2021unified}.  For circle walking, the subject population had $N = 9$ (6 male, 3 female), mass 73.6 $\pm$ 10 kg, height 1.74 $\pm$ 0.13 m, age 22.6 $\pm$ 1.7 years (mean $\pm$ s.d.). For short distance walking, the subject population had $N = 10$  (9 male, 1 female), body mass 72.1 $\pm$ 13.1 kg, height 1.69 $\pm$ 0.11 m, and age 25.5 $\pm$ 3 years.

% Results and Discussion can be combined.
\section*{Results}
\begin{figure*}[h!]
\begin{center}
 \includegraphics{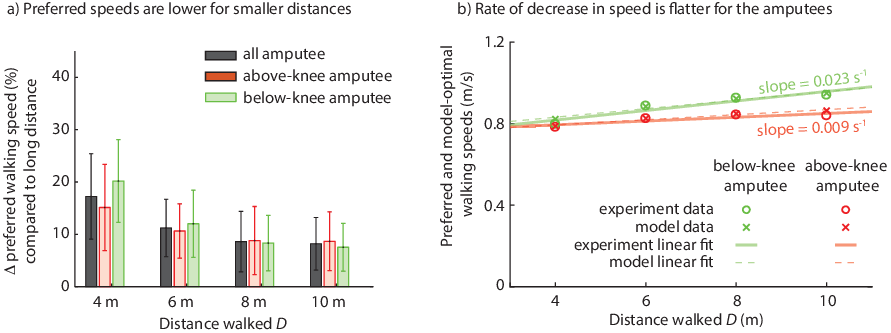}
	\caption{\textbf{Decrease in preferred walking speed with distance walked for amputees.} a) Amputees showed a decrease in average preferred walking speed for short distances. b) The rate of change in preferred walking speed with distance for the unilateral amputees via linear fits.}
	\label{fig:JFoot2}
\end{center}
\end{figure*}

\paragraph{Preferred walking speeds are lower for shorter distances.}
Subjects showed a decrease in the average preferred walking speed for shorter distances. Across all subjects, the preferred walking speed for each of the short distances (4\,m, 6\,m, 8\,m and 10\,m) was significantly lower than the preferred walking speed for the long-distance 23 m trial (each $p<10^{-4}$ using bootstrap and $p\le 3 \times 10^{-3}$ with left sided t-tests, all with Bonferroni corrections). The percentage decreases in the amputees' preferred walking speeds, compared to the long distance 23 m trial, are shown in Fig \ref{fig:JFoot2}a. On average, the percentage decrease in speed is larger for shorter distances: across all 12 amputee subjects, the percentage decreases ranged from 8.2\% $\pm$ 5\% for the 10 m walk to 17.2\% $\pm$ 8\% for the 4 m walk (mean $\pm$ s.d.).  In above-knee amputees, the percentage decreases ranged from 8.7\% $\pm$ 5.6\% for the 10 m walk to 15.1\% $\pm$ 8\% for the 4 m walk (mean $\pm$ s.d.). In below knee amputees, the percentage decreases ranged from 7.5\% $\pm$ 4.5\% for the 10 m walk to 20\% $\pm$ 8\% for the 4 m walk. 

Fitting a straight line to speed versus distance data for the short distance walking bouts, we find that the speed increases with distance with positive slope 0.015 s$^{-1}$ ($p = 3 \times 10^{-5}$, fraction variance explained: adjusted $R^2 = 0.92$). This confirms systematic increase in speed with distance.  Separating out the subjects into above and below knee amputees suggests a faster speed-increase with distance for below-knee amputees  (Fig \ref{fig:JFoot2}b), but we do not perform formal statistical comparison due to low subjects numbers.

\begin{figure*}[h!]
\begin{center}	
 \includegraphics[scale=0.97]{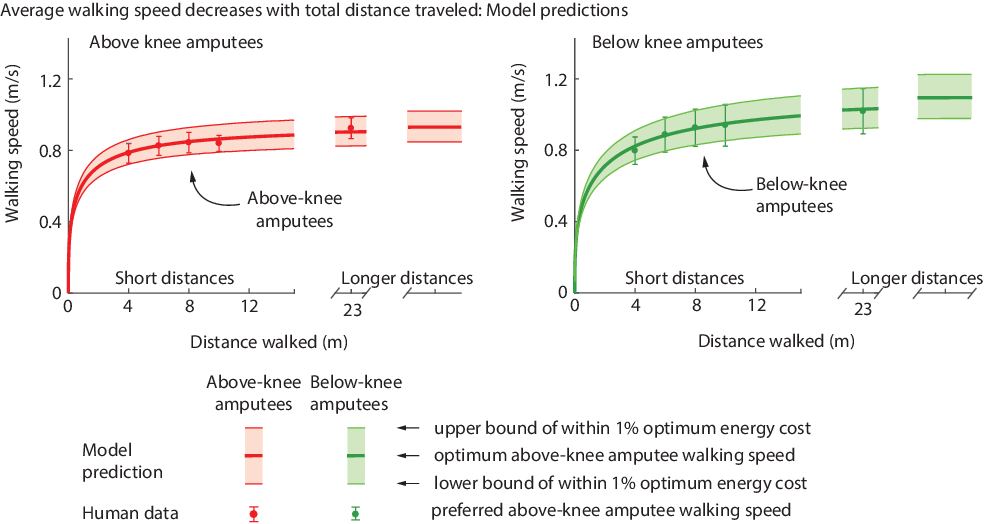}
	\caption{\textbf{Minimization of total metabolic cost captures slower short-distance walking speeds.} The total cost of the walking a short distance includes a term due to constant-speed cost and a changing-speed cost. We find that minimizing this total cost predicts the observed trends in changing preferred walking speed with distance for both amputees and non-amputees. The error bars for human data represent standard errors, and the filled bands represent the set of all speeds within 1\% of the energy optimal energy cost. The changing speed cost was obtained via inverse optimization, but the qualitative trends remain as long as the cost is positive.}
	\label{fig:JFoot3}
\end{center}
\end{figure*}

\paragraph{Increased cost of walking and changing speeds is consistent with speed-distance relationships.}
Our simple optimization-based model of short distance walking predicts the distance-dependence of optimal walking speed for both above and below knee amputees (Fig \ref{fig:JFoot2}-\ref{fig:JFoot3}), when we take into account the increased constant-speed metabolic cost of walking previously measured in experiments \cite{Jaegers1993,Genin2008} and an increased cost of changing speeds compared to young non-amputee subjects \cite{seethapathi2015metabolic}. Using inverse optimization, we estimated the scaling factor $\lambda$ for the cost of changing speeds to be 2.47 for below-knee amputees and 2.50 for above-knee amputees to best explain the observed walking speeds (Fig \ref{fig:JFoot3}a-b). With these $\lambda$ values, the mean amputee preferred walking speeds plus one standard error is within 1\% of the optimal energy costs from this model (Fig \ref{fig:JFoot3}). 

\begin{figure*}[h!]
\begin{center}
 \includegraphics[scale=0.98]{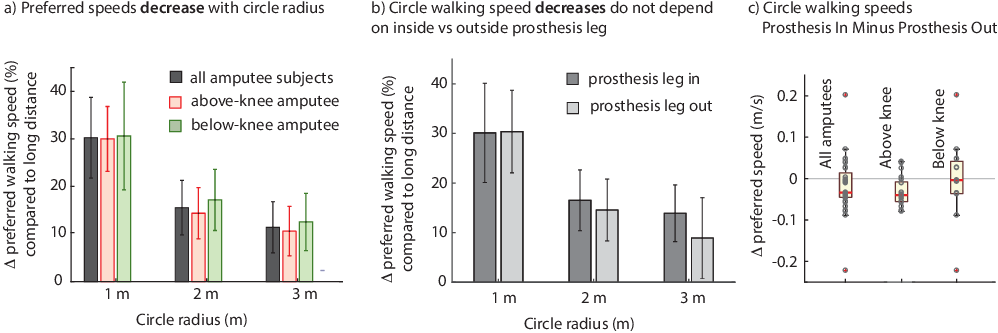}
	\caption{\textbf{Preferred walking speeds for circle walking.} a) The preferred walking speed for all the unilateral amputees showed a decrease with radius of the circle walked. b) Amputees, when pooled together, did not show a significant difference in preferred walking speed when walking with the prosthesis-leg inside versus outside the circle. c) Above knee amputees show a greater walking speed on average when the prosthesis leg is outside the circle.}
	\label{fig:JFoot4}
\end{center}
\end{figure*}

\paragraph{Preferred walking speeds are lower for smaller radii.}
Subjects showed a decrease in preferred walking speeds in smaller circles. Pooled across all subjects, the preferred walking speed for each of the radii (1 m, 2 m, 3 m) was smaller than for the 23 m straight-line walking (each $p<10^{-4}$ using bootstrap and $p <10^{-3}$ using left sided t-tests, all with Bonferroni correction).  Indeed, irrespective of type of amputation, all the unilateral amputees walked slower on circles on \textit{every trial} compared to their straight-line walking trials (Fig \ref{fig:JFoot4}a), explaining the low $p$ values.

\begin{figure*}[h!]
\begin{center}
 \includegraphics[scale=0.98]{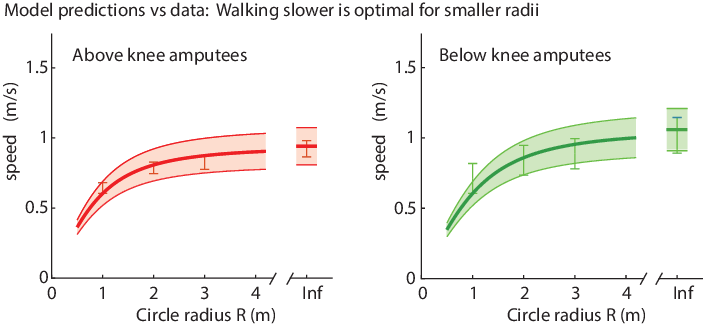}
	\caption{\textbf{Optimal walking speeds for circle walking.} Minimizing the energy cost of walking in a circle predicts slower walking for smaller circles. Error bars shown for the data correspond to one standard error about the mean, and these are generally within the set of all speeds within 1\% of the optimal energy costs (the shaded bands shown, as in Fig \ref{fig:JFoot3}. The cost of turning was obtained via inverse optimization, but the general monotonic trend between speed and radius will remain as long as the cost is positive. }
	\label{fig:JFoot5}
\end{center}
\end{figure*}

Fitting a straight line between speed and radius with subject specific offsets, we get a slope of 0.09 s$^{-1}$ ($p < 10^{-5}$ and adjusted fraction variance explained, $R^2$ = 0.92). Using inverse optimization to fit this speed dependence on circle radius, we find that the best-fit scaling coefficient $b_2$ for the cost of turning was about 6 Watts/kg/(s$^{-1}$)$^2$ more than non-amputees from \cite{brown2021unified} for below-knee amputees and about 8 Watts/kg/(s$^{-1}$)$^2$ for above-knee amputees. The fit model contains the observed walking behavior inside the band of speeds that are within 1\% of the optimal costs  (Fig \ref{fig:JFoot5}).

% add remarks about gait initiation and termination, about the cost of force rates, and about how such issues may act in addition to 

\paragraph{Preferred walking speeds for amputees walking in circles is dependent on turning direction.}
We had subjects walk both clockwise and anti-clockwise along circles drawn on the ground. We did this so as to check for any effects due to having the prosthesis-leg as the pivot, as opposed to the intact leg as the pivot. Considering all amputees together, we did not find significant differences between the two conditions ($p = 0.4$, figure \ref{fig:JFoot4}b-c). However, as a post hoc exploratory test, just considering the above-knee amputees with trials for all radii pooled  , we found that they walked slightly faster when the prosthetic leg was outside the circle ($v_\mathrm{prosthesis-out}-v_\mathrm{prosthesis-in} = 0.03 \pm 0.035$ ms$^{-1}$, $p<10^{-3}$; see figure \ref{fig:JFoot4}c).

\paragraph{Preferred gait initiation swing is usually with the affected limb.} As another post hoc exploratory analysis, we noted whether the subjects stepped forward with their affected or unaffected limb for their very first step. Stepping forward with the affected limb corresponds to the first swing phase being with the affected limb and the first stance phase being with the unaffected limb. We found that 9 out of 12 subjects had over 80\% of their first steps be with their prosthetic foot; the other three subjects had 69\%, 36\%, and 0\% of their steps start with swinging the affected limb. These leading limb preferences for gait initiation are similar to those found in \cite{vrieling2008gait}.

\section*{Discussion}
Preferred walking speed is often used as a measure of progress in walking rehabilitation for various populations, for instance, persons with neuromuscular disorders and amputees \cite{batten2019gait}, with faster speeds being considered better. Past theoretical and experimental work on non-amputee subjects shows that the speed at which people choose to move depends on the constraints of the motion itself, for instance, the distance walked \cite{seethapathi2015metabolic} and the curvature of the motion \cite{brown2021unified}, with people moving at a lower speed than physically possible due to some other objective, such as minimizing energy. Here, we find that these observations extend to amputee populations as well. In the paragraphs below, we discuss implications of these findings below.

We find that unilateral amputees slow down on average when walking shorter distances, as do non-amputees \cite{seethapathi2015metabolic}. This implies that the distance over which the preferred walking speed is measured and interpreted during rehabilitation may systematically overestimate or underestimate the progress that the patient has made. Studies involving walking rehab sometimes measure the preferred walking speed over short distances e.g.,\cite{graham2008assessing,peters2013assessing,wilson2013utilization}. In order to circumvent distance-effects on speed, we suggest measuring the speed over a few distances, not just one or two. Also, when comparing the preferred speed values for amputees to the non-amputee values, we suggest comparing to the values for the same distance walked; see Supplementary Appendix S1, Figs S3-S4 for a visualization of non-amputee data from \cite{seethapathi2015metabolic} overlaid on Figs \ref{fig:JFoot2}-\ref{fig:JFoot3} here.

Everyday walking consists of not just straight line walking but also turning.  Here, we find that unilateral amputees also slow down when taking sharper turns (circles of smaller radii) similar to non-amputees \cite{brown2021unified}. When comparing the turning speeds of amputees with those of non-amputees, it may be useful to make the comparison at fixed radii to factor out this effect of radii; see Supplementary Appendix S1, Figs S3-S4 for a visualization of non-amputee data from \cite{brown2021unified} overlaid on Figs \ref{fig:JFoot4}-\ref{fig:JFoot5} here.  For these reasons, we propose that circle-walking at multiple fixed radii lends itself as an additional useful measure of turning performance during rehabilitation.
% Unlike the straight line walking trials, we were not interested in capturing the effects of speeding up from and slowing down to rest for circle walking. We believe that the multiple laps used makes such distance-dependence effects negligible.

Most studies on energy optimality in locomotion involve constant-speed straight line walking on treadmills. Here, we provide evidence that energy-minimization can predict aspects of non-steady or non-straight-line overground walking behavior in an amputee population. The qualitative model predictions of slower speeds for smaller distances and slower speeds for smaller radii are true as long as the energy cost of changing speeds or turning is positive \cite{seethapathi2015metabolic,brown2021unified}.

Fitting the short-distance walking model and circle-walking model to the amputee walking speed behavior, we found that the scaling factors for the cost of changing speeds and for the cost of turning ($\lambda$ and $b_2$, respectively in Eqs \ref{eq:changingshortdist} and \ref{eq:Eturning} respectively) are much higher than for non-amputees. Specifically, the scaling factor $\lambda$ for the cost of changing speeds was around 2.5 for the amputee populations here compared to about 0.67 for a younger non-amputee population in \ref{seethapathi2015metabolic}; these $\lambda$ values were different by a factor of nearly 3.9. Similarly the estimated coefficient $b_2$ for the cost of turning was six to eight times larger for amputees compared to non-amputees (around 6 and 8 for amputees in this study, compared to 0.966 for non-amputees in \cite{brown2021unified}). Thus, mechanistically, these values suggest much larger costs for gait initiation, termination, and turning for amputees compared to the younger non-amputees in \cite{seethapathi2015metabolic,brown2021unified}. To test whether these increase cost estimates are entirely due to walking with a prostheses or if they may be confounded by age, we must repeat the experiments in age-matched non-amputees, while also controlling for any other confounders. To test whether these increased cost estimates are indeed real, one could directly measure the metabolic cost of changing speed and the cost of turning, as in \cite{seethapathi2015metabolic,brown2021unified}. Stability considerations may be an alternative to metabolic cost being the determinant of reduced speeds, especially for turning in a circle. One could examine this alternative hypothesis by having subjects use different speeds and estimating simple measures of stability \cite{wang2014stepping,seethapathi2019step}.

We measured speed by explicitly performing preferred speed experiments. An alternative to such measurements during prescribed tasks is to track subjects' speeds and movements all day using body-worn sensors such as pedometers, IMUs, GPS, etc \cite{dudek2008ambulation,rebula2013measurement,hordacre2014use}. Ultimately, it is these speeds during daily living that is of relevance to quantifying mobility. Such ambulatory measurements provide an opportunity to independently corroborate the results in this study, by characterizing the speeds over bouts of different lengths and walks with turns that naturally occur during daily life.

We have studied the preferred walking speeds of unilateral amputees wearing a particular passive prosthetic leg, namely, the Jaipur Foot prosthesis, used in a number of developing countries. Using preferred walking speeds as a performance measure may be more relevant where the resources available are limited, where access to other measures of performance, such as using a gait lab with motion capture and force plates, may be limited. Thus, we feel that our conclusions regarding the quantification of preferred walking speeds as a performance measure would be more relevant where walking-speed-based mobility measures would be more exclusively used. % TO DO: make this more prominent.

For short-distance walking, we measured the average speed over the whole bout of the amputees. This average speed includes the acceleration and deceleration periods \cite{miff2005temporal}. So the reduction in average speed is partly due to a greater portion of the bout being spent in acceleration-deceleration and partly due to reduced steady walking speed. The presence of smooth acceleration-deceleration phases may also suggest an additional energy cost penalty for speed change rates, not accounted for here. The acceleration-deceleration phase was included in computing the mean speed in the earlier non-amputee study as well \cite{seethapathi2015metabolic}. In that study, walking slower for shorter distances was a real choice, as the subjects could certainly cover the distance in shorter time. Similarly, prior studies with amputee populations found that they can typically walk much faster than their preferred speeds (e.g., \cite{ralston1958energy,Jaegers1993}). 

The relatively small sample size of subjects was sufficient for the specific hypotheses we tested, given the within-subjects comparisons and repeated measures design of our study. A follow-up study will be conducted to investigate the population with a larger sample size. A larger and more diverse sample will also allow us to observe the effect of other covariates such as body mass, age, gender, amputation level, height, time spent each day on their feet, and years since amputation. It would also be useful to repeat these experiments in other subject populations, including amputees wearing other prostheses with different mechanical properties.

In conclusion, we have demonstrated that walking speed depends on the distance traveled for straight line walking and the turn radius while turning in amputee subjects wearing the Jaipur foot prostheses. Given the relative simplicity of these tasks, we propose that distance-dependence of walking speeds and radius-dependence of walking speeds be used as measures of mobility not just in amputees, but also other subject populations such as the elderly and those with or recovering from other movement disorders.

\begin{table}
\scriptsize
\begin{center}
\begin{tabular}{||m{10mm} m{10mm} m{10mm} m{10mm} m{8mm} m{8mm} m{10mm} m{10mm} m{10mm} m{10mm} m{15mm} ||} 
 \hline
Subject index & Age (years) & Mass (kg) & Height (m) & $\ell_\mathrm{pros}$ (m) & $\ell_\mathrm{intact}$  (m) & BK/AK & Leg with prosthesis & Gender  & $T_\mathrm{onLegs}$ (hours)  & Years since
amputation  \\  [0.5ex] 
 \hline\hline
1&    21&    81&    1.72&    0.95&    0.99&    BK&    L&    M&    3&    5    \\
2&    40&    55&    1.54&    0.99&    0.96&    AK&    L&    M&    10&    18    \\
3 &    50 &    90 &    1.84 &    1.07 &    1.05 &    AK &    L &   M &    10 &    30    \\
4 &    35 &    66 &    1.77 &    0.98 &    0.99 &    AK &    L &    M &    10 &    29    \\
5 &    24 &    80 &    1.55 &    0.86 &    0.86 &    AK &    R &    F &    5 &    14    \\
6 &    38 &    60 &    1.71 &    1.00 &    0.99 &    AK &    L &    M &    3 &    8    \\
7 &    40 &    70 &    1.64 &    0.99 &    0.99 &    BK &    L &    M &    3 &    17    \\
8 &    53 &    64 &    1.63 &    1.02 &    1.02 &    AK &    R &    M &    1 &    20    \\
9 &    60 &    65 &    1.65 &    0.99 &    0.99 &    BK &    L&    M &    2 &    23    \\
10 &    22 &    47 &    1.67 &    1.05 &    1.05 &    AK &    L &    M &    4 &    2    \\
11 &    59 &    57 &    1.56 &    0.96 &    0.96 &    BK &    L &    M &    1 &    9    \\
12 &    26 &    54 &    1.74 &    1.01 &    1.00 &    BK &    R &    M &    10 &    5 \\ [1ex] 
 \hline
\end{tabular}
\end{center}
\caption{Information regarding subjects. The height is in shoes. The lengths $\ell_\mathrm{pros}$ and $\ell_\mathrm{intact}$ are, respectively, leg lengths  on the prosthesis and intact sides. The column with BK or AK indicates whether the subject had below-knee or above-knee amputation, here, considered equivalent to trans-tibial or trans-femoral amputation. The leg with prosthesis column indicates whether the left (L) or the right (R) leg had the prosthesis. The duration $T_\mathrm{onLegs}$ indicates a self-reported time duration that the subject spends each day on their legs, standing or walking. }
\label{table:Subjects}
\end{table}

\normalsize

\section*{Supporting information}
%
%% Include only the SI item label in the paragraph heading. Use the \nameref{label} command to cite SI items in the text.
%\paragraph*{S1 Fig.}
%\label{S1_Fig}
%{\bf Bold the title sentence.} Add descriptive text after the title of the item (optional).
%
%\paragraph*{S2 Fig.}
%\label{S2_Fig}
%{\bf Lorem ipsum.} Maecenas convallis mauris sit amet sem ultrices gravida. Etiam eget sapien nibh. Sed ac ipsum eget enim egestas ullamcorper nec euismod ligula. Curabitur fringilla pulvinar lectus consectetur pellentesque.
%
\paragraph*{S1 Data.}
% \label{S1_Data}
% {\bf Contains all individual-level raw data.}  
Compressed data file contains subject properties and the durations of walking bouts both for straight line and circle walking trials. 
%
%\paragraph*{S1 Video.}
%\label{S1_Video}
%{\bf Lorem ipsum.}  Maecenas convallis mauris sit amet sem ultrices gravida. Etiam eget sapien nibh. Sed ac ipsum eget enim egestas ullamcorper nec euismod ligula. Curabitur fringilla pulvinar lectus consectetur pellentesque.
%
\paragraph*{S1 Appendix.}
% \label{S1_Appendix}
Figs S1-S4 in this Supplementary Appendix reproduces Figs \ref{fig:JFoot2}-\ref{fig:JFoot3} here, with non-amputee data from \cite{{seethapathi2015metabolic},brown2021unified} re-drawn and overlaid for a qualitative visualization, and putting the results in this manuscript in the context of the older results. 

%
%\paragraph*{S1 Table.}
%\label{S1_Table}
%{\bf Lorem ipsum.} Maecenas convallis mauris sit amet sem ultrices gravida. Etiam eget sapien nibh. Sed ac ipsum eget enim egestas ullamcorper nec euismod ligula. Curabitur fringilla pulvinar lectus consectetur pellentesque.

\section*{Acknowledgments}
The authors thank Dr. Harlal Singh Mali for hosting N.S. in his lab briefly during her visit to Jaipur and for initially facilitating interactions. NS was funded by a Schlumberger fellowship grant and Ohio State University travel and research awards. MS was supported by NSF CMMI grant no. 1254842.

 % \nolinenumbers

% \bibliographystyle{plos2015} % plos2015 has already been declared earlier in the preamble
% \bibliographystyle{vancouver} % I think they may prefer Vancouver style
% \bibliography{jaipur}

% Either type in your references using
% \begin{thebibliography}{}
% Text
% \end{thebibliography}
%
% or
%
% Compile your BiBTeX database using our plos2015.bst
% style file and paste the contents of your .bbl file
% here. See http://journals.plos.org/plosone/s/latex for 

% 
%\begin{thebibliography}{10}
%
%\bibitem{bib1}
%Conant GC, Wolfe KH.
%\newblock {{T}urning a hobby into a job: how duplicated genes find new
%  functions}.
%\newblock Nat Rev Genet. 2008 Dec;9(12):938--950.
%
%\bibitem{bib2}
%Ohno S.
%\newblock Evolution by gene duplication.
%\newblock London: George Alien \& Unwin Ltd. Berlin, Heidelberg and New York:
%  Springer-Verlag.; 1970.
%
%\bibitem{bib3}
%Magwire MM, Bayer F, Webster CL, Cao C, Jiggins FM.
%\newblock {{S}uccessive increases in the resistance of {D}rosophila to viral
%  infection through a transposon insertion followed by a {D}uplication}.
%\newblock PLoS Genet. 2011 Oct;7(10):e1002337.
%
%\end{thebibliography}

% \clearpage

\end{document}